%% file: main.tex
\pdfoutput=1
\RequirePackage[unicode,bookmarksnumbered]{hyperref} 

 \documentclass[manuscript,nonacm]{acmart}
\usepackage{minted}
\usepackage{array}
\usepackage{multirow}
\usepackage{multicol}
\usepackage{booktabs}
\usepackage{tabularx}

\usepackage[compatibility=false]{caption}
\usepackage{hyperxmp}
\AtBeginDocument{%
  \providecommand\BibTeX{{%
    \normalfont B\kern-0.5em{\scshape i\kern-0.25em b}\kern-0.8em\TeX}}}

\graphicspath{ {./endless_total/} }
\graphicspath{ {./endless_not_total/} }

\begin{document}

\title{Theorem Provers: One Size Fits All?}

\author{Harrison Oates}
\affiliation{%
  \institution{The Australian National University}
  \streetaddress{52 Daley Road}
  \city{Acton}
  \state{ACT}
  \country{Australia}
  \postcode{2601}
}
\email{u7468212@anu.edu.au}

\author{Hyeonggeun Yun}
\affiliation{%
  \institution{The Australian National University}
  \streetaddress{52 Daley Road}
  \city{Acton}
  \state{ACT}
  \country{Australia}
  \postcode{2601}
}
\email{geun.yun@anu.edu.au}

\author{Nikhila Gurusinghe}
\affiliation{%
  \institution{The Australian National University}
  \streetaddress{52 Daley Road}
  \city{Acton}
  \state{ACT}
  \country{Australia}
  \postcode{2601}
}
\email{u7300256@anu.edu.au}

\renewcommand{\shortauthors}{Oates, Yun and Gurusinghe}

\begin{abstract}
  Theorem provers are important tools for people working in formal verification. There are a myriad of interactive systems available today, with varying features and approaches motivating their development.
  These design choices impact their usability, alongside the problem domain in which they are employed.
  We test-drive two such provers --- Coq and Idris2 --- by proving the correctness of insertion sort, before providing a qualitative evaluation of their performance. We then compare their community and library support.
  This work helps users to make an informed choice of system, and highlight approaches in other systems that developers might find useful.

\end{abstract}



\received{22 September 2023}
\received[revised]{3 November 2023}

\maketitle

\section{Introduction}
The field of theorem proving systems seems to be at a turning point, with the role of a prover undergoing an evolution from a mere assistant to a necessity. Distinguished Mathematics Professor Kontorovich  mentions this in their foreword to the Experimental Mathematics’ Special Issue on Interactive Theorem Provers \cite{Kontorovich2022}, positing that in the near future highly regarded journals will only accept proofs formalized in theorem proving systems. Going further, Kontorovich states that this is not as radical as it may seem since similar transitions have occurred: there was a point in the 19th century when it became required for calculus research papers to present proofs in Cauchy’s $\epsilon$/$\delta$ formalism. The necessity of theorem proving systems is further underscored by \citeauthor{nawaz2019survey} \cite{nawaz2019survey}, who points to incidents such the 2009 crash of Air France Flight 447 (228 deaths), the 2009 Washington Metro collision (80 severe injuries, 9 deaths); and the infamous Therac 25 incidents, resulting in several peoples' deaths.
These incidents all arose from faulty software and could have been avoided if these systems were validated mathematically; demonstrating why formalisation with theorem provers are a necessity for mission critical systems \cite{nawaz2019survey}. Consequently, the importance of mathematical formalization within the literature justifies a broader understanding of the state of the theorem proving landscape. Comparison studies, especially those focusing on novel theorem proving systems, can adequately provide this within the literature. We make a specific contribution to this domain through comparing the interactive provers Coq and Idris2 for their approach through a user-experience lens.

\section{Background}
\subsection{The provers}
Coq, first released in 1989, is an interactive theorem prover (ITP) built upon the Calculus of Inductive Constructions \cite{Coq23}, a constructive type system, and developed in OCaml. The prover allows users to state theorems and software specifications and then interactively develop formal proofs of these theorems through tactics. Generated proofs are machine-checked by a certification kernel to verify the validity of the developed proof. Some notable uses of Coq include the development of CompCert, a formally-verified optimizing compiler for C99 \cite{Leroy2009}, as well as a proof of the four colour theorem \cite{Gonthier2008}. 


Released in 2020, Idris2 is a relatively new purely functional programming language descended from Idris/Idris1 (initial release 2007) with first class dependent types \cite{edwinQTT}. In contrast to Coq, it is primarily designed as a general-purpose programming language rather than as a dedicated theorem prover.
Like Coq, Idris2 uses a constructive logic \cite{idris2Type}, but with extensions from quantitative type theory that allow the user to specify when a function can run, not just a function's intended, proven purpose.
The utility of this is immediately apparent, with the developers giving the example of guaranteeing an ATM only dispenses cash at most once per card insertion  \cite{edwinQTT}.

\subsection{Related work}

There is no shortage of research comparing theorem provers on the attributes and features associated with them. 
Many of these are in-depth with expansive investigation such as \cite{Weidijk03}, which offers a fifteen-way comparison of industrial-strength mathematical provers, aiming to demonstrate the variety of approaches available for the problem of formalizing mathematics.
These studies tend to focus on the size of the standard library, the strength of their logic, and the level of automation that each system offers.
\cite{Weidijk03} was a companion paper to \cite{17Provers2006}, which compiled proofs of the irrationality of \(\sqrt{2}\) in a superset of these provers as a means of direct comparison.
The latter paper's focus was on the appearance of proof formalization results, rather than on the formalization experience of users.
Other papers, such as \cite{nawaz2019survey}, do not test-drive the provers and instead rely on surveys from developers and online benchmarks to draw their conclusions, although these are fewer in number.

However, there is a shortage of research comparing novel theorem provers. This shortage is especially evident for newer systems.
Examples include \cite{yushkovskiy2018comparison}, \cite{Zammit1997}, and \cite{Jakubiec1997}, which all evaluate and compare Coq and Isabelle/HOL --– the regularity of this topic is emblematic of a common trend within the theorem proving comparison literature.
Interestingly, this disparity seems to occur independently of the market share  of a theorem proving system as for instance Lean --- the theorem prover of choice for many studies aiming to formalize mathematical concepts --- seems to have very little coverage in terms of studies directly comparing it to other theorem provers.
This is because many of the in-depth studies tend to focus on popular and more well-established theorem provers such as Coq or Isabelle/HOL — provers that were released over thirty years ago. 
As a result, there is a distinct scarcity of literature that compares newer systems such as Idris2 against these well-established theorem provers. 
Moreover, papers may be rendered obsolete due to their age and subsequent development of the examined systems.  

As much as this could be justified by how novel these theorem provers are, it is worrying that, in many cases, there is little literature comparing these theorem provers at all. As a result, it is difficult to ascertain the true dimensions of the current theorem proving landscape in terms of what options exist, the power of these options as well as the features that they offer. Zammit mentions this in their comparison of Coq and HOL \cite{Zammit1997}, stating that many users of theorem provers lack the perspective of knowing more than one such system, justifying this by stating that becoming proficient with a theorem proving system is a time-consuming task. Going further Zammit argues that comparison studies which compare a user’s preferred theorem prover with others concretes their understanding of the system they are familiar with whilst providing them better perspective about others. Hence, these types of studies form an important portion of the literature and can be informative in terms of how one is to go about proving the correctness of  a given algorithm.

\section{Comparing the provers}

\subsection{Installing the provers}
The experience of setting up the provers varied greatly between the systems. 
Coq provides pre-built binaries for all operating systems, alongside package offerings for the main Linux package managers, while Idris2 requires compilation from source.
We downloaded Coq from \texttt{archlinux-community} and interacted with the system through the VSCoq extension for VS Code\footnote{\url{https://github.com/coq-community/vscoq}}.
Proof General\footnote{\url{https://proofgeneral.github.io/}} or CoqIDE\footnote{\url{https://opam.ocaml.org/packages/coqide/}} are more commonly used than VSCoq according to a 2022 community survey \cite{CoqCommunitySurvey2022}, however we chose the alternate tooling to better reflect how a new user, coming from Haskell, might start learning the system.

Idris2 notionally supports all major operating systems, requiring either Chez Scheme or Racket in order to bootstrap Idris2, with Windows additionally requiring MSYS2 to simulate a unix-like environment.
When we attempted installation on Windows, we ran into difficulties with Chez Scheme's path involving spaces which could not be resolved even after help from the language's community forums.
We eventually switched to installing the system via WSL, which required only a single \texttt{make} command after downloading the source.
Like Coq, we used a VSCode extension\footnote{\url{https://github.com/bamboo/idris2-lsp-vscode}} to install the language server protocol for quality-of-life improvements.

\subsection{Proof writing}
To provide a benchmark by which to compare the experience, we showed the correctness of polymorphic insertion sort in each prover using multisets. In Haskell:
\begin{minted}{haskell}
    insert :: Ord a => a -> [a] -> [a]
    insert i []       = []
    insert i (x : xs) = if i <= x then (i : x : xs) else (x : insert i xs)
    
    sort :: Ord a => [a] -> [a]
    sort []       = []
    sort (x : xs) = insert x (sort xs)
\end{minted}

To do this, we require a predicate \texttt{sorted}, defined inductively as follows:
\begin{align}
    &\text{sorted } [] \\
    &\forall n, (\text{sorted [n]}) \\ 
    &\forall m n p, (\text{leq }n\, m \to \text{sorted (}m : p) \to \text{sorted (}n : m : p))
\end{align}
alongside a total ordering \texttt{leq} and a function to return the contents of a list,
\[\text{contents} : \text{list } A \to \text{multiset } A.\]
Multisets \(A\) and \(B\) are considered equal if and only if the multiplicities of all elements in \(A\) and \(B\) are the same.

We need to prove three lemmas (explicitly or otherwise):
\begin{align}
     \forall x : A, l : \text{list } A, (\text{contents (insert }x\, l) &=  \text{contents (}x : l)) \\
     \forall x : A, l : \text{list } A, (\text{sorted }l &\to \text{sorted (insert }x\, l)) \\
     \forall l : \text{list } A, (\text{contents }l &= \text{contents (sort }l))
\end{align}

In order to prove the main theorem:
\[
    \forall l : \text{list } A, (\text{contents } l = \text{contents (sort }l) \land \text{sorted (sort }l)).
\]

We were permitted to use the standard libraries of each language as required. The proofs can be be found at \url{https://github.com/HarrisonOates/COMP2560-Theorem-Prover-Code} and we refer to them throughout this section.


\subsubsection{Defining functions} 

The Haskell specification translates to each language relatively easily, with some language specific quirks.
\begin{minted}{coq}
Inductive sorted : list A -> Prop := 
  | sorted_nil : sorted []
  | sorted_singleton : forall n, sorted [n]
  | sorted_cons : forall n m p, leq n m -> sorted (m :: p) -> sorted (n :: m :: p).
\end{minted}
The Coq specification for \texttt{sorted} uses the type \texttt{Prop} of propositions, but if-statements require a statement of type \texttt{bool}. The types are not interchangeable because booleans are guaranteed to be decidable while propositions are not, so this necessitated a decidability hypothesis \texttt{leq\_dec}. Hypotheses for the transitivity and totality of  \texttt{leq} also need to be specified:
\begin{minted}{coq}
Hypothesis leq_dec : forall a b, {leq a b} + {~ leq a b}.
Hypothesis leq_trans : forall a b c, leq a b -> leq b c -> leq a c.
Hypothesis leq_total : forall a b, ~ leq a b -> leq b a.
\end{minted}
Hence \texttt{leq\_dec} is used to write insertion sort:
\begin{minted}{coq}
Fixpoint insert (i : A) (l : list A) :=
    match l with
    | [] => [i]
    | x :: xs => if leq_dec i x then (i :: x :: xs) else (x :: insert i xs)
    end.
    
Fixpoint sort (l : list A) : list A :=
    match l with
    | [] => []
    | x :: xs => insert x (sort xs)
    end.
\end{minted}
When a program invokes the created proof for an arbitrary type, the user will need to provide an implementation of \texttt{leq} for that type and a proof that the stated hypotheses are satisfied.

Unlike Coq, in Idris2 there is no dedicated terminology for different parts of the proof, such as \texttt{Inductive}, \texttt{Hypothesis} and \texttt{Fixpoint}, as they are all defined as their own functions. Each proposition is constructed as types to form functions, as can be seen in type \texttt{IsSorted} that defines the \texttt{sorted} predicate.
\begin{minted}{idris}
data IsSorted : (xs:Vect n e) -> Type where
    IsSortedZero :
        IsSorted Nil
    IsSortedOne  :
        Ord e =>
        (x:e) ->
        IsSorted (x::Nil)
    IsSortedMany :
        Ord e => 
        (x:e) -> (y:e) -> (ys:Vect n'' e) ->  -- (n'' == (n - 2))
        (IsLte x y) -> IsSorted (y::ys) ->
        IsSorted (x::(y::ys))
\end{minted}
With this implementation, any \texttt{Vect} in \texttt{IsSorted} type now guarantees that it is sorted in non-decreasing order. This type is then used in process of deriving and proving the predicates of \texttt{insert}. It should also be noted that the data type as a whole is used to define the function but its specific cases, such as \texttt{IsSortedZero} and \texttt{IsSortedOne}, are utilized for the implementation.
\begin{minted}{idris}
insert :
    Ord e =>
    (xs:Vect n e) -> (y:e) -> (IsSorted xs) -> 
    (xs':(Vect (S n) e) ** ((IsSorted xs'), (ElemsAreSame (y::xs) xs')))
insert Nil y IsSortedZero =
    ([y] ** (IsSortedOne y, XsIsXs [y]))
insert (x::xs) y proofXXsIsSorted =
    let proofHeadOfXXsIsX = the (HeadIs (x::xs) x) MkHeadIs in
    case (insert' (x::xs) y proofXXsIsSorted proofHeadOfXXsIsX) of
        (xs' ** (proofXsNewIsSorted, proofHeadXsNewIsXOrY, proofYXXsIsXsNew)) =>
            (xs' ** (proofXsNewIsSorted, proofYXXsIsXsNew))
\end{minted}
A quirk of the implementation is that the proof uses the \texttt{Vect} type rather than \texttt{List}, which seems counterintuitive.
\texttt{Vect}, as the type of ``lists with length'' \cite{IdrisDocumentation} was chosen for the implementation over \texttt{List} based on an experienced user suggestion. 
We note that \texttt{Data.Vect} provides \texttt{fromList} and \texttt{Foldable} provides \texttt{toList}, which both preserve ordering of the elements. Accordingly, \texttt{List} is also supported by this implementation.
In the REPL, this detail is unimportant (because the size of the list is known at runtime) and sort can be directly applied without repercussions.

Such types can be used as a part of theorems, asking for the proof for its property in order for the theorems (functions) to be compiled, as can be seen in the main theorem \texttt{insertionSort} (lines 180-185). This idea of having `propositions as types' provides a unified approach between computational and proof-oriented constructs \cite{idrisTypes}. In other words, there are no specific terminologies insinuating the purpose of functions like \texttt{variable} and \texttt{Hypothesis}, nor requiring separate variables like \texttt{Prop} for ``things to be proven" and \texttt{bool} for ``things to be computed", in contrast to Coq.

Specifying \texttt{sort} from the Haskell definition was trivial in Coq, with a clear correspondence between the two functions (lines 36-46). Idris2, however, builds the proof directly into the function, so the relation between our Haskell \texttt{sort} and Idris2' \texttt{insertionSort} (lines 182-195) is less clear. This represents a major difference in philosophy: In Coq, the user creates the function and then proves its properties, while Idris2 creates and proves a function correct simultaneously.

\subsubsection{Interacting with the engine} 
The main way of manipulating the proof state in Coq is through tactics, which constitute steps in the proof. Each tactic is stepped through and checked by the kernel in real time, throwing an error in the IDE if it is not a valid application of a tactic. Coq also accepts inserting proof terms into the system directly in a manner similar to Idris2, which we discuss below. This offers finer-grained control for users that desire it.


Idris1 used to have tactics in a similar vein to Coq \cite{IdrisDocumentation}. The majority of tactics were deprecated during the transition to Idris2, as this version focused on providing a more minimal core for better efficiency and dependability.
Nevertheless, due to the Curry-Howard Correspondence,
the following relationships between proof system and Idris2's computation model can be established \cite{Curry}:
\begin{itemize}
    \item Propositions in Logic corresponds to Type in Idris2
    \item Proofs in Logic corresponds to Terms/Programs in Idris2
    \item Simplification of proofs in Logic corresponds to evaluation of program/normalisation of proofs in Idris2
\end{itemize}
In summary, Idris2 accepts an implementation of function to be a valid proof, where the function is the theorem consisting of propositions expressed as types, if it compiles without any holes. Holes (\texttt{LHS = ?\{hole\_name\}}) are unwritten right hand sides of proofs that can be revised later when writing functions incrementally.
\\

\subsubsection{Recovery and Maintainability} 


Recovery refers to the ability to change approach while writing a proof. Both Coq and Idris2 have features that facilitate recovery as many `interactive' theorem provers innately do. Coq can undo a tactic or series of tactics by stepping backwards through proof state, so navigating to an earlier instance of the current proof is simple. If the user wishes to add a theorem above the current proof, the system allows for their proof to be temporarily omitted with the \texttt{Admitted} command. This accepts the stated proposition as an axiom, with the intention that the user returns to implement the proof later, similar to holes in Idris2.
In a similar manner, using helper functions in Idris2 allows for incremental changes so logical or syntactic errors can be isolated to each function.
The REPL helps facilitate this process by catching errors at compile time.

Maintainability is facilitated by both systems in the form of proof searches. In Idris2, this occurs through the REPL command \texttt{:ps n f} or simply hovering the cursor on the hole, which searches for a proof to fill hole \texttt{f} at line \texttt{n}. The example below clearly demonstrates how the right hand side of \texttt{insertionSort}'s base case can be immediately filled with help from such interactive feature:

\begin{minted}{idris}
186 insertionSort Nil = ?insertionSort_rhs_0
-- (terminal command)
InsertionSort> :ps 186 insertionSort_rhs_0
([] ** (IsSortedZero, NilIsNil))
\end{minted}

Coq takes a slightly different approach to search, instead specifying the `shape' of a theorem as a line in the proof:
\begin{minted}{coq}
    Search (_ + _ <= _ + _). 
    (* Results in a list of all imported theorems that match the shape, e.g *)
    Plus.plus_le_compat_r_stt: forall n m p : nat, n <= m -> n + p <= m + p 
    Nat.crossmul_le_addsquare:
      forall a b : nat, 0 <= a -> 0 <= b -> b * a + a * b <= a * a + b * b
\end{minted}
Coq's approach has the advantage of the user not needing to change to change keyboard focus in order to find their results.
Search in both systems helps the user to use previous results, thereby avoiding duplication of code.



\subsection{Running programs}
The method by which to compile and run programs varies across the systems. Within a proof session, Coq can evaluate terms, printing an output when the checker steps over the step:

\begin{minted}{coq}
    Definition simpleReverseList := rev [1;2;3].
    Eval compute in simpleReverseList. (* = [3;2;1] : list nat *)
\end{minted}
Needless to say, this isn't particularly useful. Instead, programs written in Coq can be `extracted' to Haskell, OCaml, or Scheme, allowing the user to compile and run it in their preferred language. 
This is natively supported through the \texttt{Extraction} keyword.
Proofs of the correctness of this process are discussed in \cite{Letouzey2008}.

This conversion does not come without its disadvantages.
When we ran small-scale benchmarks of the extracted Haskell for a list of \texttt{Int}s, the extracted version took 23\% more time and 13\% more memory than our hand-written insertion sort for a 1000-element reversed order list, with the performance penalty decreasing to 9.2\% more time and 7\% more memory for a 10000-element list.
The conversion did not use native Haskell types, instead creating its own \texttt{List} type, and required instantiation of \texttt{leq\_dec} and type \texttt{A} before it would run. 
While there exist techniques for using native types at time of extraction to improve performance, such as extracting \texttt{nat} to \texttt{Int}, the size of \texttt{Int} is finite. Accordingly, using this technique there exist theorems that can be proven in Coq that would not hold in the resulting Haskell extraction \cite{Appel:SF3}, impacting guarantees of correctness.

This additional layer of complexity makes running extracted Coq programs efficently a more involved process than Idris2, which can compile from source to an executable without additional steps.
The snippet below corroborates the easiness to run the final \texttt{insertionSort} that guarantees the propert of \texttt{sorted} and preserves the \texttt{contents}:
\begin{minted}{idris}
insertionSort :
    Ord e =>
    (xs:Vect n e) ->
    (xs':Vect n e ** (IsSorted xs', ElemsAreSame xs xs'))

-- (terminal command)
InsertionSort> insertionSort [2,1,3]
([1, 2, 3] ** (sorted proof, contents proof (too long to be displayed)))
\end{minted}

Such readiness to be computed is thought to be a significant advantage for programmers who require proof of their executable functions' properties. One very important aspect of Idris2 during run time is the Quantitative Type Theory (QTT) \cite{edwinQTT}. Every single variable in Idris2 has one of three quantities that corresponds to the number of its usage at run time, which are either 0, 1 or unrestricted. Having the quantity as 0, also known as erasure, brings preciseness to examining which values are relevant at run time, and which are compile time only. Some of the common erasure variables are type of element \texttt{e} and size of \texttt{Vect} \texttt{n}, as they are only needed to initialize function properties at compile time. The linearity with quantity of 1 indicates that a variable must be used exactly once, whether the type variable is pattern matched or the function variable is applied. Although not compulsory, users can inspect the condition of holes easier
by explicitly specifying such quantity. Exploitation of the QTT also improves performance as irrelevant variables at run time are automatically handled.

\subsubsection{Termination}
Consider the following non-terminating Haskell program:

 \begin{minted}{haskell}
     endless :: Int -> Int
     endless x = endless (x + 1)
 \end{minted}
This will compile and run fine. Writing an equivalent program in Coq, however, and we run into trouble.
Coq requires a proof of termination for each defined function. In the case of Coq, if this is not provided the totality checker will throw an error. Furthermore, the proof can be found automatically through \texttt{Program Fixpoint} in some instances, otherwise the prover will help the user to prove the termination obligation using a strictly decreasing measure. 
Idris2, on the other hand, does not require termination by default, and will readily accept the above program unless the flag \texttt{-}\texttt{-total} is passed to the interpreter. Each function can also go under totality checking with \texttt{:total \{function\}}. By not forcing termination, Idris2 users can sacrifice correctness to write functions where termination would not be desirable, such as the main loop of a web server. This is a clear benefit over Coq, which would require non-terminating code to be written in the target extraction language. 


\subsection{Community/Library Support}
Community support for a theorem prover is important, as it is the community that develops many of the systems' libraries and help new users get up to speed with the system. As the exact size of a community is difficult to ascertain and are spread across many online spaces, we use research mentions according to the ACM digital library\footnote{\url{dl.acm.org}} and the number of GitHub contributors as a proxy. Since Idris2 is a relatively new system, ``Idris" was searched instead. Using ``Idris2" as a keyword results in 3 papers.

We summarize these findings in Table \ref{tab:communitySize} below.  Coq has the clear advantage in terms of both GitHub contributors and amount of related research, with the amount of related research for 
Idris2 being significantly smaller. This is unsurprising, considering Coq has well over thirty years of continuous use and development behind it. The number of GitHub contributors for Coq, nearly six times more than Idris2, further highlights its sizable community. 

\begin{table}[h!]
  \caption{Dedicated communities and their sizes as at October 13, 2023}
  \label{tab:communitySize}
  \begin{tabular}{ccc}
    \toprule
    Prover  & Github Contributors & Related Research count\\
    \midrule
    Coq    & 230 & 2345 \\
    Idris2 & 40 & 439 (3)\\
   
  \bottomrule
\end{tabular}
\end{table}

In order to take advantage of the community, a theorem prover needs substantive, easy-to-use libraries.
Coq's official package index formalizes a myriad of mathematics for use by other projects, easily accessible through OPAM \cite{Coq23}.
Idris2 has a package manager in development \cite{Idris2Pack} and a canonical list of current libraries on their Github page, but this does not yet match Coq in expansiveness or ease of use. This issue will lessen with time.

\section{Future work}
There are a number of potential avenues for future investigation. 
First, expanding the methodology of this paper to other theorem provers such as Isabelle, Lean or HOL4 would help to build a more comprehensive survey of the current theorem-proving landscape. This would build directly upon this paper's dual contributions, both expanding information for users to make an informed choice of system and highlighting features that developers might find useful in maintaining their current system.
Second, we propose an in-depth comparison of what is achievable in terms of program performance once proofs are compiled. This could be further extended to compare the specific mechanisms behind each theorem prover's compiler.
Motivated by our initial benchmarks of code generated by Coq extraction, we finally suggest investigating optimization methods to prepare Coq programs for extraction while maintaining correctness. Reducing the performance penalty of verified code would be a significant contribution because it makes the adoption of formal methods more palatable to industry, which in turn would improve safety in critical systems.


\section{Conclusion}
In conclusion, we have shown that Coq and Idris2 each have their own distinguishing features that makes them more suited towards specific proof styles.
We provide a summary in Table   ~\ref{tab:overall}.
\begin{table}[h!]
    \centering
    \caption{Overall comparison between the theorem provers}
    \label{tab:overall}
    \begin{tabular}{p{0.14\linewidth} p{0.37\linewidth} p{0.38\linewidth}}
        \toprule
        Factor & Coq   & Idris2 \\
        \midrule
        Installation Difficulty & Simple - binary + source available  & Linux: easy, Windows: much harder \\ 
        \midrule
        Maintainability & Strong type checking with some \;inferencing ability, offers search \;capabilities  & Strong type checking allows safer \;incremental changes with assist from holes\\ 
        \midrule
        Interactiveness with prover & Tactic-based manipulation with \;option for proof-term manipulation  & Proof-term manipulation with many features to assist a user from adding clauses, case splitting to proof searching \\ 
        \midrule
        Readability & Reads like pen + paper proofs. Haskell-like syntax  & Very readable to those who have \;experience with Haskell, but does not resemble a traditional proof \\ 
        \midrule
        Ideal scenario & Users wanting to take advantage of significant library + community \;support and a more conventional  style of proof  & If a user desires executable code and is familiar with functional programming\\
        \bottomrule
    \end{tabular}
\end{table}

\begin{acks}
To Dirk Pattinson, our supervisor, for your time and energy in helping us succeed. 
We additionally thank Donovan Crichton for sharing their expertise in Idris2.
\end{acks}

\pagebreak

\bibliographystyle{ACM-Reference-Format}
\input{sample-base.bbl}

\end{document}